\documentstyle[12pt]{article}
\newcommand{\ptmis}{{ {\rm p} \hspace{-0.53 em} \raisebox{-0.27 ex} {/}_T }}
\newcommand{\beq}{\begin{equation}}
\newcommand{\eeq}{\end{equation}}
\def\lsim{\raise0.3ex\hbox{$\;<$\kern-0.75em\raise-1.1ex\hbox{$\sim\;$}}}
\def\gsim{\raise0.3ex\hbox{$\;>$\kern-0.75em\raise-1.1ex\hbox{$\sim\;$}}}
\def\VEV#1{\left\langle #1\right\rangle}

\def\np#1#2#3{           {\it Nucl. Phys. }{\bf #1} (19#2) #3}
\def\pl#1#2#3{           {\it Phys. Lett. }{\bf #1} (19#2) #3}
\def\pr#1#2#3{           {\it Phys. Rev. }{\bf #1} (19#2) #3}
\def\prep#1#2#3{         {\it Phys. Rep. }{\bf #1} (19#2) #3}
\def\prl#1#2#3{          {\it Phys. Rev. Lett. }{\bf #1} (19#2) #3}
\def\etal{\hbox{\it et al., }}

\begin{document}

\begin{titlepage}
\rightline{FTUV/93-50   \hskip 0.9cm IFIC/93-31}
\rightline{IFT-P.072/93 \hskip 0.5cm IFUSP-P 1084}
\rightline{MAD/PH/810}
\noindent
\begin{center}
{\large \bf Searching for an Invisibly Decaying Higgs Boson \\
in $e^+e^-$, $e\gamma$, and $\gamma\gamma$ Collisions}
\vskip 0.5cm
{\bf O.\ J.\ P.\ \'Eboli$^1$} and {\bf M.\ C.\ Gonzalez-Garcia$^2$}\\
{Physics Dept., University of Wisconsin, \\
Madison, WI 53706, USA}\\
\vskip 0.3cm
{\bf A.\ Lopez-Fernandez$^3$}\\
{PPE Division, CERN, CH-1211 Geneve 23, Switzerland }\\
\vskip 0.3cm
{\bf S.\ F.\ Novaes$^4$}\\
{Instituto de F\'{\i}sica Te\'orica, Universidade Estadual Paulista,\\
Rua Pamplona, 145 -- CEP 01405-900, S\~ao Paulo, Brazil}\\
\vskip 0.3cm
{\bf J.\ W.\ F.\ Valle$^5$}\\
{ Instituto de F\'{\i}sica Corpuscular - C.S.I.C.\\
Dept.\ de F\'{\i}sica Te\`orica, Universitat de Val\`encia\\
46100 Burjassot, Val\`encia, Spain}
\vskip0.3cm
{\bf ABSTRACT}\\
\end{center}

Higgs bosons can have a substantial ``invisible'' branching ratio in
many extensions of the Standard Model, such as  models  where the Higgs
bosons  decay predominantly into light or massless weakly interacting
Goldstone bosons. In this work, we examine the production mechanisms and
backgrounds for invisibly decaying Higgs bosons at the Next Linear
$e^+e^-$ Collider operating in the modes $e^+e^-$, $e\gamma$, and
$\gamma\gamma$. We demonstrate that such machine is much more efficient
to survey for invisibly decaying Higgs bosons than the  Large Hadron
Collider at CERN.
\end{titlepage}


\setcounter{page}{1}

\section{Introduction}
\label{intro}

One of the primary goals of the Next Linear Collider (NLC) is to
unravel the nature of the symmetry breaking mechanism of the
electroweak interaction. Recently, there has been a great deal of
interest in the study of Higgs boson signatures at such a machine
\cite{nlc2}. In this work, we analyze the signal and backgrounds
at the NLC for the production of Higgs bosons that decay
invisibly, {\em i.e.\/} into very weakly interacting particles.
We study the three possible modes of operation of the NLC,
$e^+e^-$, $e\gamma$, and $\gamma\gamma$, assuming that the it
will operate with  a center-of-mass energy of $\sqrt{s}=500$ GeV
and a luminosity ${\cal L}_{ee} = 10^4$ pb$^{-1}$/year. Our
results show that the best mode of operation to search for
invisibly decaying Higgs bosons is the $e^+e^-$ one, where  Higgs
bosons  with masses up to $200$ GeV can be observed provided
their coupling to the $Z$ is larger than $1/3$ of its value in
the Standard Model.  Fully coupled Higgs bosons can be detected
up to masses of $300$ GeV.  Therefore, the mass range of
``invisible'' Higgs boson that can be probed at NLC is larger
than the corresponding one for LEPII \cite{alfonso,inv:lep} and even for
the Large Hadron Collider at CERN (LHC) \cite{inv:had}.

There are many reasons to speculate that there exist additional
Higgs bosons in nature, besides the one predicted by the Standard
Model. For instance, the extension of the minimal Standard Model
provided by supersymmetry and the desire to tackle the hierarchy
problem \cite{revsusy} predicts a larger physical Higgs spectrum.
Another interesting motivation for the enlargement of the Higgs
sector is to generate the observed baryon excess by electroweak
physics \cite{Kuz}. Although there is not yet a consensus on this
matter, there are claims that suggest that a successful
electroweak baryogenesis imposes, in the Standard Model, a
stringent limit on the higgs mass, namely $m_h \lsim 40$ GeV
\cite{Dineetal92a}.  However, this bound is in conflict with the
constraint placed by the direct search of the standard model
Higgs boson  at the LEP experiments around the $Z$ peak
\cite{LEP1}, {\it i.e.\/} $m_{H_{SM}} \gsim 60$ GeV.  A way to
reconcile the LEP data and baryogenesis is to consider models
with new Higgs bosons \cite{2Higgs}. Moreover, such additional
Higgs bosons could be intimately related to the question of
neutrino masses \cite{BGLAST}.  In fact, one specially attractive
motivation for extending the Higgs sector is the generation of
neutrino masses \cite{fae}, whose existence is hinted by present
data on solar and atmospheric neutrinos, as well as cosmological
observations related to the large scale structure of the universe
and the possible need for hot dark matter.

Amongst the extensions of the Standard Model, which have been
suggested to generate neutrino masses, the so-called majoron
models are particularly interesting and have been widely
discussed \cite{fae}. The majoron is the Goldstone boson
associated with the spontaneous breaking of lepton number.
Astrophysical arguments based on stellar cooling rates constrain
its couplings to charged fermions \cite{KIM}, while the LEP
measurements of the invisible $Z$ width substantially restrict
the majoron couplings to gauge bosons. In particular, models
where the majoron is not a singlet under the $SU(2)_L \times
U_Y(1)$ symmetry \cite{GR} are now excluded if lepton number is
violated only spontaneously \cite{LEP1}. There is, however, a
wide class of models  motivated by neutrino physics
\cite{JoshipuraValle92} which is characterized by the spontaneous
violation of a global $U(1)$ lepton-number symmetry by a singlet
vacuum expectation value. Unlike the original model of this type
\cite{CMP}, this new class of models may naturally explain the
neutrino masses required by astrophysical and cosmological
observations without introducing any high mass scale. Another
example of this type is provided by supersymmetric extensions of
the Standard Model where R parity is spontaneously violated
\cite{MASI}.

In any of these models with the spontaneous violation of a global
$U(1)$ symmetry around or below the weak scale, the corresponding
Goldstone boson can have significant couplings to the Higgs
bosons, even though its couplings to matter are suppressed. This
implies that the Higgs boson can decay, with a substantial
branching ratio, into the invisible mode, $h \to J\;+\;J$, where
$J$ denotes the majoron \cite{JoshipuraValle92,Joshi92,HJJ}.

An invisible decay of the Higgs boson leads to events with large
missing energy.  However, in order to have an observable
signature, it  must be produced in association with another
particle, such as $Z$, $W$, or $t$, which is used to tag the
event. The production of  an invisibly decaying Higgs boson was
previously  considered at LEP in association with a $Z$
\cite{inv:lep}  and at hadron colliders in association with a $Z$
or a $t\bar{t}$ pair \cite{inv:had}. In this paper we concentrate
on the possibility of identifying an invisibly decaying Higgs
boson at the three different modes of the NLC.

An important feature of this next generation of linear $e^+e^-$
colliders is that they should also be able to operate in the $e\gamma$
or $\gamma\gamma$ modes. The conversion of electrons into photons can
occur via the laser backscattering mechanism \cite{las0}, which leads to
a photon beam with almost the same energy and luminosity of the parent
electron beam. This makes the NLC a very versatile machine that could
employ energetic electrons and/or photons as initial states.

The outline of the paper is as follows. In Sec.\ \ref{toy} we briefly
review the features of a simple model exhibiting invisibly decaying
Higgs bosons.  Section \ref{ee} is devoted to the study of the $e^+e^-$
mode of the collider, whereas the modes $e\gamma$ and $\gamma\gamma$
are discussed in Sec.\ \ref{gam}. We summarize our conclusions in Sec.\
\ref{dis}.


\section{The Simplest Model}
\label{toy}

In order to illustrate the main features of invisibly decaying Higgs
bosons, let us analyze a simple model that contains the Standard Model
scalar Higgs doublet plus an additional complex singlet $\sigma$, which
exhibits a nonzero vacuum expectation value $\VEV{\sigma}$ responsible
for breaking  a global $U(1)$ symmetry. The scalar potential of this
model is \cite{JoshipuraValle92,Joshi92}
\begin{equation}
\label{V1}
V = \mu_{\phi}^2\phi^{\dagger}\phi
+\mu_{\sigma}^2\sigma^{\dagger}\sigma
+ \lambda_{1}(\phi^{\dagger}\phi)^2+
 \lambda_2
(\sigma^{\dagger}\sigma)^2
+\delta (\phi^{\dagger}\phi)(\sigma^{\dagger}\sigma) \; .
\end{equation}
Terms like $\sigma^2$ were omitted in Eq.\ (\ref{V1}) since we required
this model to exhibit a $U(1)$ invariance under which $\sigma$
transforms nontrivially and $\phi$ trivially. Let
\[
\sigma
\equiv \frac{w} {\sqrt{2}} + \frac{(R_2+iI_2)}{\sqrt{2}}
 \;\;\; \hbox{and} \;\;\;
\phi \equiv \frac{v}{\sqrt 2}+ \frac{(R_1+iI_1)}{\sqrt {2}} \; ,
\]
where we have set $\VEV{\sigma} = w/\sqrt{2}$ and $\VEV{\phi} =
v/\sqrt{2}$. The potential (\ref{V1}) leads to a physical massless Goldstone
boson, namely the majoron $J \equiv {\rm Im}\; \sigma$, and two neutral
CP even mass eigenstates $H_i$ ($i$= 1,2)
\beq
H_i={\hat O}_{ij}\;R_j \; ,
\eeq
where ${\hat O}_{ij}$ is an orthogonal mixing matrix with mixing angle
$\theta$ given, in terms of the potential parameters and the
vacuum expectation values, by
\beq
\tan 2\theta = \frac{\delta v w}{\lambda_1 v^2- \lambda_2 w^2} \; .
\eeq

In this simple model only the doublet Higgs boson $\phi$ couples to $Z$,
$W$, and charged fermions in the weak basis. As a result, the CP even
states $H_i$ interact with these particles only through their $R_1$
component.  After diagonalizing the scalar boson mass matrix, one finds
that the two CP even mass eigenstates $H_i$ possess the following
couplings:
\begin{equation}
\begin{array} {l}
\label{HZZ1}
{\cal L}_{H_iZZ}
=(\sqrt 2 G_F)^{1/2}~ M_Z^2~ {\hat O}_{i1}~ Z_{\mu}Z^{\mu} H_i \; , \\
{\cal L}_{H_iWW}
=2(\sqrt 2 G_F)^{1/2}~ M_W^2~ {\hat O}_{i1}~ W^+_{\mu}W^{-\mu} H_i \; , \\
{\cal L}_{H_if\bar f}
=(\sqrt 2 G_F)^{1/2}~ m_f~ {\hat O}_{i1}~ \bar f f H_i \; .\\
\end{array}
\end{equation}
As we can see, the strength of the $H_i$couplings are reduced, in
relation to the Standard Model, by a factor $\epsilon_i =
\hat{O}_{i1}$. Therefore, as long as the mixing appearing in Eq.\
(\ref{HZZ1}) is ${\cal O}(1)$, all CP even Higgs bosons can have
significant production rates, similar to the ones of the Standard
Model. In general, the $SU(2)$ custodial symmetry implies that
the Higgs couplings to $Z$ and $W$ bosons are suppressed by the
same factor, which can, in principle, be different from the
suppression factor of the couplings to fermions.

We now turn to the Higgs boson decay rates, which are sensitive to the
details of the mass spectrum and thus to the Higgs potential.  For
definiteness, we focus on the model given in Eq.\ (\ref{V1}) in which
case the invisible width of the $H_i$ is
\begin{equation}
\label{HJJ}
\Gamma(H_i\rightarrow JJ)=\frac{\sqrt 2 G_F}{32 \pi} M_{H_i}^3 g^2_{H_i JJ} \;
,
\end{equation}
where the corresponding couplings are $g_{H_iJJ}=\tan\beta \; {\hat
O}_{i2}$, with $\tan\beta = v/w$. The width for $H_i\rightarrow b
\overline b$ gets diluted compared to the Standard Model one,
because of mixing effects. Explicitly one has
\begin{equation}
\Gamma(H_i\rightarrow b \overline b)=\frac{3\sqrt 2 G_F}{8
\pi}M_Hm_b^2
\left (1-4 \frac{m_b^2}{M_H^2} \right)^{3/2}g^2_{H_ib\overline b} \; ,
\end{equation}
which is smaller than the Standard Model prediction by the factor
$g_{H_ib\overline b}^2$, where $g_{H_ib\overline b}= {\hat O}_{i1}$.
Heavy Higgs bosons can also decay into $ZZ$ or $W^+W^-$ pairs, however,
the partial widths into these modes are diminished by the same factor
appearing in the $b\bar{b}$ mode, as can be seen from Eq.\ (\ref{HZZ1}).

In summary, the branching ratio of the Higgs decay into the invisible mode
$JJ$ depends upon the mixing angles $\beta$ and $\theta$ and for a large
fraction of the parameter space the invisible Higgs decay mode is expected to
have quite large branching ratio \cite{alfonso}. This is characteristic of
models where there exists a global symmetry that is broken around the weak
scale.

{}From the existing $Z$ sample at LEP \cite{alfonso} it is possible to
obtain limits on invisibly decaying Higgs bosons. Moreover, it is
possible to set Higgs boson mass limits that are not vitiated by
detailed assumptions on its mode of decay. In this paper we stress that
this invisible decay can be used as a useful signature for Higgs bosons
at linear $e^+e^-$ colliders.


\section{Invisible Higgs Bosons at the NLC: the $e^+e^-$ Mode}
\label{ee}

At the NLC, the main production mechanisms for $H_i$ involve their
couplings to heavy particles such as $Z$ and $W$ bosons and top quarks,
depending on the collider mode chosen.  For the NLC operating in the
$e^+e^-$ mode, the most important reactions for Higgs production
\cite{hunterguide} are  the Higgs bremsstrahlung off a $Z$ boson
\beq
e^+e^-\rightarrow Z^* \rightarrow Z H
\label{prod}
\eeq
and the $WW$ fusion process
\beq
e^+e^-\rightarrow \nu\bar\nu W^* W^* \rightarrow \nu\bar\nu H  \; .
\eeq
For Higgs bosons decaying invisibly, this second mechanism becomes
irrelevant since it leads to an undetectable final state. In Fig.\ 1 we
plot the cross section for the Higgs bremsstrahlung process as a
function of the Higgs mass for $\sqrt{s}=500$ GeV, assuming that the
coupling of the Higgs boson to the Z is the one predicted by the
Standard Model.

The main sources of background for invisibly decaying Higgs
bosons are
\[
\begin{array}{llll}
e^+e^- \rightarrow \nu \bar\nu Z \:\:& (\sigma=0.48 \: \mbox{pb}) & & [A]
\\[0.3cm]
e^+e^- \rightarrow W^+ W^-  \:\: & (\sigma=7.8 \: \mbox{pb})
 &  \rightarrow  (q\bar q') \:[\ell] \:\nu  &[B] \\
& & \rightarrow  (\ell \bar \ell)  \:      \nu  &[C]\\[0.3cm]
e^+e^- \rightarrow e\nu_e W  \:\: & (\sigma=6.0 \: \mbox{pb})
  & \rightarrow   (q\bar q') \:[e] \:\nu_e  &[D]\\
&  & \rightarrow   (e^+e^-) \: \nu_e  &[E] \\
\end{array}
\]
where the particles in square brackets escape undetected and the fermion
pairs in parentheses have an invariant mass close to the $Z$ mass.  The
equations above show the total cross sections for each process
without taking into account any kinematical cut or branching ratio.

The Higgs bremsstrahlung mechanism (\ref{prod}) with the $Z$
decaying into hadronic modes possesses large backgrounds due to
the processes [$B$] and [$D$], even when we demand the $Z$
invariant mass reconstruction.  For this reason we will consider
in our study only the leptonic decay modes  $Z\rightarrow e^+e^-$
and $Z\rightarrow \mu^+\mu^-$.  In this case, the signature for
an invisibly decaying Higgs boson will be two charged leptons
with invariant mass compatible with the $Z$ mass, plus missing
transverse momentum.  The requirement of missing transverse
momentum eliminates further backgrounds like $\gamma \gamma$ and $\ell
\bar{\ell}(\gamma)$ events, and makes the background  coming from the
process [$E$] to be  irrelevant.  Finally, the most dangerous
background that we are left with is process [$A$] \cite{mele}. In order
to suppress this contribution,  we impose a further cut on the
reconstruction of the $Z$ energy
\beq
E_Z(m_H)=\frac{s+m_Z^2-m_H^2}{2\sqrt{s}}~ \pm~ \Delta E \; .
\eeq
where $\Delta E$ is the size of the uncertainties expected for
the energy measurement at NLC, which we assume to be $\Delta
E = 10$ GeV.  This cut relies upon the fact that the energy of
the $Z$ is fixed when it is produced in association with the
Higgs boson.

Further suppression of this background can be attained  using the
fact that the $Z$'s in the signal are produced at larger polar
angles than the ones in the background, as can be seen in Fig.\
2. Therefore, we impose an additional angular cut $|\cos\theta_Z|
< 0.7$.  After all these cuts, the background [$C$] becomes very
small as shown in Fig.\ 3. This figure also shows the number
of events we are left with for the signal and backgrounds [$A$]
and [$C$] for $\epsilon^2\times$ BR$_{\hbox{invis}}=1$, where
BR$_{\hbox{invis}}$ is the invisible branching ratio of the Higgs
boson,  and an integrated luminosity of $10^4$ pb$^{-1}$.

In Fig.\ 4 we show the 95\% confidence level discovery contours in the
$\epsilon^2 \times$ BR$_{\hbox{invis}}$ versus $M_H$ plane that can be
explored at the NLC in the $e^+e^-$ mode. From this figure, we can learn
that invisibly decaying Higgs bosons with masses below $200$ GeV can be
detected provided their coupling to the $Z$ is higher than 1/3 of its
value in the Standard Model, while fully coupled Higgs bosons can be
discovered up to masses of almost $300$ GeV.


\section{Invisible Higgs Bosons at NLC: the $\gamma\gamma$
and $e\gamma$ Modes} \label{gam}

In a linear collider, it is possible to transform an electron beam into
a photon one through the process of laser backscattering \cite{las0}.
This mechanism relies on the fact that Compton scattering of energetic
electrons by soft laser photons gives rise to high energy
photons, that are collimated in the direction of the incident electron.
For unpolarized initial electrons and laser, the spectrum of
backscattered photons is \cite{laser}
\begin{equation}
F_L (x) \equiv \frac{1}{\sigma_c} \frac{d\sigma_c}{dx} =
\frac{1}{D(\xi)} \left[ 1 - x + \frac{1}{1-x} - \frac{4x}{\xi (1-x)} +
\frac{4
x^2}{\xi^2 (1-x)^2}  \right] \; ,
\label{f:l}
\end{equation}
with
\beq
D(\xi) = \left(1 - \frac{4}{\xi} - \frac{8}{\xi^2}  \right) \ln (1 + \xi) +
\frac{1}{2} + \frac{8}{\xi} - \frac{1}{2(1 + \xi)^2} \; ,
\eeq
where $\omega_0$ is the laser energy, $\sigma_c$ is the Compton
cross section and $\xi \simeq 4 \omega_0 E /m^2$, with $m$ and
$E$ being the electron mass and energy respectively.  In contrast
to the bremsstrahlung spectrum, the backscattered photon
distribution (\ref{f:l}) peaks close to the maximum allowed value
for the photon energy, which occurs at $ x_{\mbox{max}}=\xi/(1+\xi)$.
Moreover, the luminosity of the backscattered photon
beam can be very close to the one of the parent electron beam
provided that there is no $e^+e^-$ pair creation by the
interaction of the backscattered photons with the laser. In our
calculation we have chosen $\omega_0$ such that it  maximizes the
backscattered photon energy without spoiling the luminosity
through $e^+e^-$ pair creation. This choice leads to $\xi = 2 ( 1
+ \sqrt{2})$ and the maximum of the spectrum is located at
$x_{\mbox{max}} \simeq 0.83$.

The laser backscattering mechanism transforms the NLC into a very
versatile machine that can operate in  two additional modes  which are
$e\gamma$ and $\gamma\gamma$.  The differential luminosities for these
modes can be easily obtained from Eq.\ (\ref{f:l}) and are given by
\begin{eqnarray}
\frac{d {\cal L}_{e\gamma}}{dz} &=& {\cal L}_{ee}~ 2 z   F_L(z^2) \; , \\
\frac{d {\cal L}_{\gamma\gamma}}{dz} &=& {\cal L}_{ee}~ 2z
\int_{z^2/x_{max}}^{x_{max}} \frac{dx}{x} F_L (x)F_L (z^2/x) \; ,
\end{eqnarray}
where ${\cal L}_{ee}$ is the luminosity of the NLC operating in
the $e^+e^-$ mode. The total cross section for any process in
these modes is obtained by folding the elementary cross section
with the corresponding photon luminosity
\begin{equation}
\sigma (s)=\int_{z_{min}}^{z_{max}} dz~\frac{d{\cal L}_{ij}}{dz}~
\hat\sigma_{ij} (\hat s=z^2 s)
\end{equation}
where $i,j$ is either $e\gamma$ or $\gamma\gamma$.  Here $z^2= \tau =
\hat{s}/s$, where $s$ is the total $e^+e^-$ center-of-mass energy squared
and $\hat{s}$ the $ij$ pair center-of-mass energy squared.


\subsection{The $\gamma\gamma$ Mode}

Since neutral Higgs bosons do not couple directly to photons, their
production in a $\gamma\gamma$ collider must proceed through either
higher order processes or in association with heavy particles.  The main
production mechanism for neutral Higgs bosons at such colliders is via
one-loop triangle diagrams \cite{trigg}. However, this mechanism becomes
useless for an invisibly decaying Higgs boson since it would lead to an
undetectable signature. Therefore, the most promising processes are
those in which the Higgs boson is produced in association with a
$W^+W^-$ or a $t\bar t$ pair, where the heavy particles can be used to
tag the events.
\[
\begin{array} {ll}
\gamma  \gamma \rightarrow W^+  W^-  H   & [WWH] \\
\gamma  \gamma  \rightarrow  t \bar{t}  H & [TTH]
\end{array}
\]
In the case of a standard Higgs boson, the process [$WWH$]  has been
already considered in the literature \cite{ref:wwh} and [$TTH$] was
previously analyzed in Ref.\ \cite{ref:tth,ref:ttz}. However, for the
sake of completeness, we show in Fig.\ 5 the total cross section for
these processes as a function of the Higgs mass for $\epsilon^2\times$
BR$_{\mbox{invis}}=1$ and $m_{\mbox{top} } = 140$ GeV.

The possible backgrounds for these processes come from the reactions
\[
\begin{array}{lll}
\gamma \gamma \rightarrow W^+  W^- & (\sigma=48 \mbox{ pb})&  [WW]\\
\gamma \gamma \rightarrow W^+  W^- [\gamma]
&(\sigma=1 \mbox{ pb},\;  p^\gamma_T> 20\mbox{ GeV})
& [WWG]\\
\gamma \gamma \rightarrow W^+  W^- Z \rightarrow  W^+  W^- \nu \bar \nu  &
(\sigma=4 \mbox{ fb}) &  [WWZ]\\
\gamma \gamma \rightarrow t  \bar t &  (\sigma=0.4 \mbox{ pb}) &  [TT]\\
\gamma \gamma \rightarrow t  \bar t [\gamma] &
 (\sigma= 0.2 \mbox{ fb}, \; p^\gamma_T> 20\mbox{ GeV})
& [TTG]\\
\gamma \gamma \rightarrow t \bar t Z \rightarrow  t \bar t \nu \bar \nu  &
(\sigma=0.2 \times 10^{-3} \mbox{ fb}) &[TTZ]\\
\end{array}
\]
where the $[\gamma]$ in the [$WWG$] and [$TTG$] backgrounds escapes
undetected, going into the beam pipe.  Several of these reactions have
been analyzed previously in Ref.\ \cite{ref:ttz,ref:wwz}.  In order to
ensure the reconstruction of the $W^+W^-$ and $t \bar{t}$ pairs, we
restrict ourselves to hadronic decay modes of the $W$'s and we impose
that none of the charged particles in the final state goes down the beam
pipe, that we assume to have an an angular size of $20^\circ$,
corresponding to the cut $|\cos\theta_{W,t}|<0.94$.   The potentially
larger backgrounds ([$WW$] and [$TT$]) can be eliminated by requiring
the existence of missing $p_T$ in the event, thus, we adopted the cut
$\ptmis > 40$ GeV. This requirement also discards the [$TTG$] and
[$TTZ$] backgrounds and reduces considerably the [$WWG$] one, as can be
seen in Fig.\ 6.

Unlike the $e^+e^-$ mode, it is impossible to reconstruct the invisible
invariant mass due to the unknown center-of-mass energy of the
$\gamma\gamma$ system. This constitutes an unavoidable drawback for both
the $\gamma\gamma$ and $e\gamma$ modes that might only be circumvented
by using polarized photons whose energy distribution is narrower.

In Fig.\ 6 we show, for $\epsilon^2\times$ BR$_{\hbox{invis}}=1$, the
expected number of events for the signals and backgrounds, taking into
account the hadronic branching ratio of the $W$ and the above cuts. As
can be seen from this figure, a fully coupled Higgs boson produced in
association with a $W$ pair would lead to a $2 \sigma$ signature if its
mass is lighter than $80$ GeV.  Nevertheless, this mass range would have
been almost fully explored by LEP200 \cite{alfonso} assuming that the
Higgs couplings to $Z$ and $W$ are related, as required by the $SU(2)$
custodial symmetry. For the associated production with a $t \bar{t}$
pair the cross section is very small and would lead to more than $1$
event only if the Higgs boson is lighter than $50$ GeV. For the simple
model we discussed in Sec.\ \ref{toy}, this range is already ruled out
by LEP data since the mixing angle appearing in the Higgs coupling to
fermions and to gauge bosons are the same. However, in a general model,
containing more fermions or an additional scalar doublet, the couplings
of the Higgs to gauge bosons and to fermions are independent and the
[$TTH$] process would not be directly constrained by LEP data.


\subsection{The $e\gamma$ Mode}

The main production mechanisms of Higgs bosons in a linear collider
operating in the $e\gamma$ mode are
\[
\begin{array}{ll}
e\gamma\rightarrow e\gamma\gamma\rightarrow e H & [EH]\; , \\
e\gamma \rightarrow \nu W H  & [NWH] \; , \\
e\gamma \rightarrow e Z H  & [EZH]   \; . \\
\end{array}
\]
The reaction [$EH$] was analyzed in Ref.\ \cite{eh} and it takes place
through the one-loop process $\gamma\gamma \rightarrow H$, where one of
the photons is generated via bremsstrahlung off the electron. For an
invisibly decaying Higgs this process is irrelevant since the final
electron goes down the beam pipe and there is no detectable particle in
the final state. On the other hand, the [$NWH$] reaction \cite{nwh}
leads to a $W^-$ plus missing $p_T$ signature.  However, this signal
will be buried in a huge background due to the process $e \gamma
\rightarrow W^- \nu$, which overcomes the signal by more than $3$ orders
of magnitude.  Therefore, we must restrict the search for an invisibly
decaying Higgs boson to the process [$EZH$] \cite{ezh}. In Fig.\ 7 we
show the total cross section for this process as a function of the Higgs
mass, assuming $\epsilon^2\times$ BR$_{\hbox{invis}}=1$ without applying
any cut.

The main backgrounds for this process are
\[
\begin{array} {llll}
e\gamma\rightarrow e Z &(\sigma=13 \mbox{ pb})  & &[EZ]  \; , \\
e\gamma\rightarrow e Z \nu \bar \nu & (\sigma= 1.3\times 10^{-2} \mbox{
pb})
&  & [EZNN]\; ,  \\
e\gamma\rightarrow e W^+ W^- &(\sigma= 3.8\mbox{ pb})
&\rightarrow [e] \nu e (q\bar{q}^\prime) & [EWW] \; , \\
e\gamma\rightarrow e Z \gamma &
 (\sigma=0.4 \mbox{ pb}, \;  p^\gamma_T> 15\mbox{ GeV})
&\rightarrow e Z [\gamma]  & [EZG] \; ,
\end{array}
\]
where the $[\gamma]$ in [$EZG$] escapes undetected, going into the beam
pipe, and the $(q \bar{q}^\prime)$ pair from the $W$ decays is
mistaken as a $Z$ boson. Some of these reactions have been analyzed
in references \cite{ezh,egback}. As before, the [$EZ$] background is
eliminated by demanding the event to contain missing $p_T$ ($\ptmis >20$
GeV). The bulk of the [$EWW$] events steams from effective photons
($e\gamma\rightarrow e\gamma\gamma\rightarrow e W W$), so in order to
reduce this background we require a large scattering angle for the
outgoing electron $\theta_e > 10^\circ$, that corresponds to $|\cos
\theta_e| < 0.984$.  The [$EWW$] background could be eliminated if we
restricted ourselves to leptonic $Z$ decays. However, this would also
reduce dramatically the signal, which is already small.  In our
analyses,  we consider both leptonic and hadronic decays of the $W$'s
and $Z$'s. We can further improve the signal over background ratio
noticing that in the signal the $Z$'s are more copiously produced in the
forward hemisphere, as shown Fig.\ 8, while in the [$EZNN$] process, as
well as for the $W$ in the [$EWW$] process, they are symmetrically
produced. Therefore, we also demand that  $\cos\theta_{Z,W}>0$.

In Fig.\ 9 we show the number of leptonic and hadronic events from the
signal and backgrounds for $\epsilon^2\times$ BR$_{\hbox{invis}}=1$.  As
can seen in this figure, the [$WWE$] background is still much larger
than the signal for the case of hadronic decays. Therefore, these decays
modes could not be used unless an extremely good invariant mass
resolution would permit the discrimination between hadronic decaying
$Z$'s and $W$'s. On the other hand, the signal for leptonic $Z$ decays
is very weak and only observable in a range of Higgs masses which are
already ruled out by LEP.


\section{Discussion}
\label{dis}

The Higgs boson can decay to invisible Goldstone bosons in a wide class
of models in which a global symmetry, such as lepton number, is broken
spontaneously close to the weak scale. These models are attractive from
the point of view of neutrino physics and suggest the need to search for
Higgs bosons in the invisible mode. We have investigated the reach of a
high energy linear $e^+ e^-$ collider to discover a Higgs boson decaying
invisibly.

An important feature of the next generation of linear $e^+e^-$ colliders
is that they should also be able to operate in the $e\gamma$ or
$\gamma\gamma$ modes. We have studied the possibilities of the different
modes of NLC to observe an invisible Higgs decay signature. According to
our results, it would be very difficult to observe such a invisibly
decaying Higgs boson in the $\gamma\gamma$ and $e\gamma$ modes of the
collider. In these modes, not only the cross sections for the signals
are small but also the backgrounds are difficult to be reduced  since it
is not possible to reconstruct the invisible invariant mass due to the
unknown center-of-mass energy of the $\gamma\gamma$ and $e\gamma$
systems.

The best results are obtained in the $e^+e^-$ mode of the collider.  Our
results in this case are summarized in Fig.\ 4, where we show the
exclusion contours at 95\% CL in the $\epsilon^2 \times$
BR$_{\hbox{invis}}$ vs. $m_H$ plane that can be explored at the NLC in
the $e^+e^-$ mode. Invisibly decaying Higgs bosons with masses below
$200$ GeV can be detected provided their coupling to the $Z$ is larger
than $1/3$ of the Standard Model Higgs coupling. A fully coupled Higgs
bosons can be detected up to masses of $300$ GeV. Therefore, our results
indicate that a machine like the NLC will be able to survey a much
larger $m_H$ range than the LHC.


\newpage

\begin{center}
{\bf ACKNOWLEDGMENTS}
\end{center}

O.J.P.E.\ is very grateful to the Institute of Elementary Particle
Physics Research of the Physics Department, University of
Wisconsin---Madison for its kind hospitality. This work was supported by
the University of Wisconsin Research Committee with funds granted by the
Wisconsin Alumni Research Foundation, by the U.S.\ Department of Energy
under contract No. DE-AC02-76ER00881, by the Texas National Research
Laboratory Commission under Grant No. RGFY93-221, by Conselho Nacional
de Desenvolvimento Cient\'{\i}fico e Tecnol\'ogico (CNPq/Brazil), by
Funda\c{c}\~ao de Amparo \`a Pesquisa do Estado de S\~ao Paulo
(FAPESP/Brazil), and by the National Science Foundation under Contract
INT 916182.
\vskip 2cm
\noindent
$^1$ Permanent address: Inst. de F\'{\i}sica, Universidade de S\~ao Paulo,
Caixa Postal 20516, CEP 01452-990 S\~ao Paulo, Brazil;
E-mail: EBOLI@USPIF.IF.USP.BR (InterNet) -- USPIF::EBOLI (DecNet)\\
$^2$ CONCHA@WISCPHEN (BitNet) -- 47397::CONCHA (DecNet)\\
$^3$ E-mail: ALFON@CERNVM (BitNet)\\
$^4$ E-mail: UEIFT1::NOVAES (DecNet)\\
$^5$ E-mail: VALLE@EVALUN11 (BitNet) -- 16444::VALLE (DecNet) \\


\newpage
\noindent
\begin{center}
{\bf FIGURE CAPTIONS}
\end{center}

\noindent
{\bf Fig.\ 1:} Total cross section for the process $e^+e^-\rightarrow H
Z$ as a function of Standard Model Higgs-boson mass ($m_H$) at
$\sqrt{s}=500$ GeV.

\noindent
{\bf Fig.\ 2:} Angular distribution of $Z$ in the background process
$e^+e^- \rightarrow \nu \bar\nu Z$ (dotted line) and in the Standard
Model signal $e^+e^-\rightarrow H Z$ (solid lines). The signal
distributions are shown for $m_H=20$, $200$, and $300$ GeV from the
upper to lower solid lines.

\noindent
{\bf Fig.\ 3:} Number of events for the signal $ZH$ (solid line),
assuming $\epsilon^2\times$ BR$_{\hbox{invis}}=1$, and
backgrounds $e^+e^- \rightarrow \nu \bar\nu Z$ (dashed line) and
$W^+ W^-$ (dotted line). We imposed the cut $|\cos \theta_Z| <
0.7$ and assumed an uncertainty in the $Z$ energy of $10$ GeV.

\noindent
{\bf Fig.\ 4:} Accessible region in the plane ($\epsilon^2\times$
BR$_{\hbox{invis}}$, $M_H$) at 95 \% CL. We assumed a
center-of-mass energy of $500$ GeV and an integrated luminosity
of $10^4$ pb$^{-1}$.

\noindent
{\bf Fig.\ 5:} Total cross section for the processes $\gamma \gamma
\rightarrow W^+W^- H$ (solid line) and $\gamma \gamma \rightarrow t
\bar{t} H$ (dashed line) as a function of the Higgs mass, for  $\sqrt{s}
= 500$ GeV and $m_{\hbox{top}} = 140$ GeV.

\noindent
{\bf Fig.\ 6:} Total number of [$WWH$], [$WWZ$], [$WWG$], and [$TTH$]
events in the $\gamma\gamma$ mode for $m_{\hbox{top}} = 140$ GeV, after
considering the relevant branching ratios. We imposed the cuts $|\cos
\theta_{W,t,\bar{t}} | < 0.94$, $\ptmis > 40$ GeV, and $|\cos
\theta_\gamma| > 0.94$.

\noindent {\bf Fig.\ 7:} Total cross section for the process $e\gamma
\rightarrow e Z H$ at $\sqrt{s} = 500$ GeV. We assumed  $\epsilon^2
\times$ BR$_{\hbox{invis}}=1$ and did not apply any cut.

\noindent
{\bf Fig.\ 8:} Angular distribution of the $Z$ in the background process
$e\gamma \rightarrow e Z \nu \bar\nu$ (dotted line) and in the signal
$e\gamma \rightarrow e Z H$ (solid lines). The signal distribution is
shown for $m_H = 60$ GeV.

\noindent
{\bf Fig.\ 9:} Final number of events for signal and backgrounds in the
$e\gamma$ mode. The solid lines correspond to hadronic $Z$ events while
the dashed lines correspond to the leptonic ones.

\end{document}